\documentstyle[12pt]{article}
\date{}
\newcommand{\eeq}{\end{eqnarray}}
\newcommand{\beq}{\begin{eqnarray}}
\newcommand{\bD}{{\bf D}}
\newcommand{\bE}{{\bf E}}
\newcommand{\bB}{{\bf B}}
\newcommand{\bH}{{\bf H}}
\newcommand{\br}{{\bf r}}
\newcommand{\ba}{\mbox{\boldmath $a$}}

\newcommand{\bF}{{\bf F}}
\newcommand{\bY}{{\bf Y}}
\newcommand{\bX}{{\bf X}}
\newcommand{\boX}{\overline{{\bf X}}}
\newcommand{\oQ}{\overline{{Q}}}
\newcommand{\bq}{{\bf q}}
\newcommand{\bv}{{\bf v}}
\newcommand{\bp}{{\bf p}}
\newcommand{\bx}{{\bf x}}
\newcommand{\by}{{\bf y}}

\newtheorem{TH}{Theorem}
\title{Dynamics of the Born--Infeld dyons}
\author{{Dariusz Chru\'sci\'nski\footnotemark}\hspace{4mm} and
\hspace{2mm} {Hartmann R\"omer\footnotemark}\\
       Fakult\"at f\"ur Physik, Universit\"at Freiburg\\
       Hermann-Herder-Str. 3, D-79104 Freiburg, Germany}

\begin{document}
\def\thefootnote{\relax}\footnotetext{$^*$On leave from
 Institute of Physics, Nicholas Copernicus
University, ul. Grudzi\c{a}dzka 5/7, 87-100 Toru\'n, Poland.
E-mail: dariusz@phyq1.physik.uni-freiburg.de and
darch@phys.uni.torun.pl}
\def\thefootnote{\relax}\footnotetext{$^\dagger$E-mail: 
roemer@phyq1.physik.uni-freiburg.de}
\maketitle

\begin{abstract}

The approach to the dynamics of a charged particle in the Born--Infeld
nonlinear electrodynamics developed in [Phys. Lett. A 240 (1998) 8] 
is generalized to include a Born--Infeld dyon. Both Hamiltonian and 
Lagrangian structures of many  dyons interacting with nonlinear 
electromagnetism are constructed. All results are manifestly duality 
invariant.

\end{abstract}

\vspace{4cm}

Freiburg THEP--98/9

\newpage

\section{Introduction}

In a recent paper \cite{charge} it was shown how to describe the dynamics
of a point charge in the  Born--Infeld nonlinear electrodynamics \cite{BI}
(see also \cite{IBB}). There are several motivations to study this theory
(see e.g.  \cite{charge}, \cite{IBB}, \cite{Hagiwara}, \cite{Plebanski}, 
\cite{multipole}). Moreover, it turns out that some natural objects
in string theory, so called D-branes, are described by a kind of nonlinear
Born--Infeld action \cite{string}.

The aim of the present Letter is to show that the results of \cite{charge}
could be generalized in the case when a particle carries both electric 
and magnetic charges (dyon). We do so not only for an aesthetic purpose.
It turns out that the
 Dirac idea of magnetically charged particles
\cite{Dirac} (see \cite{Goddard} for a review) has led recently to  very 
remarkable results in field and string theories (see e.g. \cite{Olive}
for an excellent introduction to this subject).

It is well known that Born--Infeld electrodynamics is duality invariant 
\cite{Gibbons} (actually, this observation was already made by Schr\"odinger
\cite{Schrodinger}). Therefore, in principle it should be possible to describe
the dynamics of dyons (like in the Maxwell case). However, it turns out
that previous attempts to the dynamics of charged particles in the
Born--Infeld theory \cite{Feenberg} (see \cite{charge} for the comparison
of results obtained in \cite{charge} and long ago by Feenberg and Pryce 
\cite{Feenberg})
are not consistent with the duality invariance of the underlying theory.
We show that the approach proposed in \cite{charge} respects this invariance.

Moreover, we present a canonical formalism for a theory describing 
the dynamics of many Born--Infeld dyons. 
Both Hamiltonian and Lagrangian structures are constructed. 
To the best of our knowledge it is the first fully
consistent canonical structure for classical electrodynamics of many 
point--like particles (including particles self--interaction). 

We stress that one usually discusses classical dyons in a different context
namely as static solutions to the non--Abelian Yang--Mills--Higgs models
(see once more \cite{Goddard}). It is conjectured that this solutions
(in the so called BPS limit \cite{BPS}) play an important role in the
nonperturbative quantum field theory \cite{Olive}. Actually, it is possible
to make a non--Abelian generalization of the Born--Infeld action
and to investigate non--Abelian Born--Infeld--Higgs models. It was shown
\cite{Nakamura} that  these models possess monopole (and dyon) solutions.
The dyons considered in the present Letter are simply puted "by hand".
However, there is a similarity between e.g. BPS mass formula for dyons
in the non--Abelian theories and a Newton--like equation (\ref{Newton})
of the present Letter. We shall comment it in the last section.

\section{Dynamical condition}

Let us briefly sketch the main result presented in \cite{charge}.
The Born--Infeld nonlinear electrodynamics \cite{BI}  is based
on the following Lagrangian (we use the Heaviside-Lorentz system of
 units with the velocity of light $c=1$):
\begin{eqnarray}    \label{Lag-BI}
{\cal L}_{BI} &:=&  \sqrt{-\mbox{det}(b\eta_{\mu\nu})}
- \sqrt{-\mbox{det}(b\eta_{\mu\nu} + F_{\mu\nu})}  \nonumber\\
&=& { b^2}\left(1- \sqrt{1-2b^{-2}S - b^{-4}P^2}\right)\ ,
\end{eqnarray}
where $\eta_{\mu\nu}$ denotes the Minkowski metric with the signature
$(-,+,+,+)$ (the theory can be formulated in a general covariant way, 
however, in this paper we will consider only the flat Minkowski space-time).
The standard Lorentz invariants
$S$ and $P$ are defined by:
$S = -\frac 14 F_{\mu\nu}F^{\mu\nu}$ and $P = -\frac 14
F_{\mu\nu}{*F}^{\mu\nu}$ (${*F}^{\mu\nu}$ denotes the dual tensor).
The arbitrary parameter ``$b$'' has a dimension of a field strength (Born
and Infeld called it the {\it absolute field}) and it measures the
nonlinearity of the theory. In the limit $b \rightarrow \infty$ the
Lagrangian ${\cal L}_{BI}$ tends to the Maxwell Lagrangian $S$.

In the presence of an electrically charged matter one usually adds
 to (\ref{Lag-BI}) the standard electromagnetic interaction term
``$j^\mu A_\mu$''. Then the complete set of field equations read:
\beq   \label{field-eqs1}
\partial_\mu {*F}^{\mu\nu} &=& 0\ ,\\  \label{field-eqs2}
\partial_\mu G^{\mu\nu} &=& - j^\nu\ ,
\eeq    
with $G^{\mu\nu} := -2
 {\partial {\cal L}_{BI}}/{\partial F_{\mu\nu}}$. Equations
(\ref{field-eqs1})--(\ref{field-eqs2}) have formally the same 
form as Maxwell equations. 
What makes the theory effectively nonlinear are the constitutive
relations, i.e. relations between inductions $(\bD,\bB)$ and
intensities $(\bE,\bH)$:
\beq
\bE(\bD,\bB) &=& \frac{1}{b^2R} \left[ (b^2 + \bB^2)\bD -
(\bD \bB)\bB \right] \ ,  \label{E}\\
\bH(\bD,\bB) &=& \frac{1}{b^2R} \left[ (b^2 + \bD^2)\bB -
(\bD \bB)\bD \right] \ ,  \label{H1}
\eeq
with $R:= \sqrt{1 + b^{-2}(\bD^2 +\bB^2) +  b^{-4}(\bD\times\bB)^2}$.

Now, let us assume that the external electric current $j^\mu$ in 
 (\ref{field-eqs2})
is produced by a point-like particle moving along the time-like trajectory
 $\zeta$. The main idea of \cite{charge} (it was developed in the
Maxwell case in \cite{KIJ}) was as follows: 
instead of solving very complicated distributional  equations
(\ref{field-eqs1})--(\ref{field-eqs2}) 
on the entire Minkowski space-time $\cal M$ let us
treat them as a boundary problem in the region ${\cal M}_\zeta :=
{\cal M} - \{\zeta\}$, i.e. outside the trajectory. Now, in order  well to
pose the problem we have to find an appropriate boundary condition 
which has to be satisfied along $\zeta$, i.e. on the boundary 
${\cal M}_\zeta$. Observe that in ${\cal M}_\zeta$ equations 
(\ref{field-eqs1})--(\ref{field-eqs2}) are homogeneous.

To find this boundary condition we have analysed the asymptotic behaviour
of the fields in the vicinity of a charged particle. The simplest way to 
do so is to use the particle's rest frame. Let $r$ denote the radial 
coordinate, i.e. a distance from a particle in its rest frame. Any vector
field $\bF=\bF(\br)$ may be formally expanded in the vicinity of a charge:
\beq
\bF(\br) &=& \sum_{n} r^n \bF_{(n)}\ , 
\eeq
where the vectors $\bF_{(n)}$  do not depend on $r$. 
The crucial observation is that the most singular part of 
$\bD$ field behaves as
\beq   \label{D-2}
\bD_{(-2)} = \frac{e{\cal A}}{4\pi}\frac{\br}{r}\ ,
\eeq
where, due to the Gauss law, the monopole part of the $r$-independent
function $\cal A$ equals 1. Note, that in the Maxwell case
${\cal A} \equiv 1$. Using (\ref{D-2}) it was shown \cite{charge}
that:
\beq   \label{chain}
\bH \sim r^{-1}\ ,\ \ \ \ 
\bE \sim r^{0}\ ,\ \ \ \ 
\bB \sim r\ .
\eeq
Moreover, the $\bE_{(0)}$ term is known explicitely:
\beq   \label{E-0}
\bE_{(0)} = \frac{be}{|e|}\frac{\br}{r}\ .
\eeq
Using these results the following theorem was proved \cite{charge}:
\begin{TH}
Any regular solution of Born--Infeld field equations with point-like
external current satisfies:
\beq  \label{E-T}
\bE_{(1)}^T = \frac {be}{4|e|}
 \left( 3\mbox{\boldmath $a$} - 
r^{-2}(\mbox{\boldmath $a$} \mbox{\boldmath $r$})
\mbox{\boldmath $r$} \right)\ ,
\eeq
where $\bE^T$ stands for the transversal part of $\bE$ and $\ba$ denotes
the particle's acceleration.
\end{TH}
According to our ``boundary philosophy'' the formula (\ref{E-T})
may be interpreted as a boundary condition for $\bE$ field on 
$\partial{\cal M}_\zeta$ and  the hyperbolicity of 
(\ref{field-eqs1})--(\ref{field-eqs2}) implies:
\begin{TH}
The mixed (initial-boundary) value problem for the Born--Infeld equations
in ${\cal M}_\zeta$  with (\ref{E-T}) playing the role of boundary
condition on $\partial{\cal M}_\zeta$ has the unique solution.
\end{TH}

The above theorem is no longer true when we consider a particle 
as a dynamical object. Choosing particle's position $\bq$ and velocity $\bv$
as the Cauchy data for the particle's dynamics let us observe that despite
the fact that the time derivatives  $(\dot{{\bD}},\dot{{\bB}},
\dot{{\bq}},\dot{{\bv}})$ of the Cauchy data are uniquely determined
by the data themselves, the evolution of the composed system is not
uniquely determined. Indeed, $\dot{{\bD}}$ and $\dot{{\bB}}$ are given
by the field equations, $\dot{{\bq}}={\bv}$ and $\dot{{\bv}}$ may
be calculated from (\ref{E-T}). Nevertheless, the initial value problem
is not well posed: keeping the same initial data, particle's trajectory 
can be modified almost at will. This is due to the fact, that now 
(\ref{E-T}) plays no longer the role of boundary condition because we
use it to as a dynamical equation to determine $\ba$. Therefore a new
boundary condition is necessary.

It was shown in \cite{charge}  that this missing condition
is implied by the conservation law of the total four--momentum for
the ``particle + field'' system: $\dot{p}^\mu =0$, where $p^\mu$
stands for the four--momentum in a fixed laboratory frame. Due to the
field equations only 3 among 4 equations are independent, i.e. the 
conservation of three--momentum 
\beq  \label{conservation}
\dot{\bp} = 0
\eeq
implies the energy conservation: $\dot{p}^0 =0$. Now, the formula 
(\ref{conservation}) is equivalent to the following Newton-like
equation:
\beq   \label{Newton}
ma_k = \frac{|e|b}{3}{\cal A}_k\ ,
\eeq
where ${\cal A}_k$ is the dipole part of $\cal A$ (see (\ref{D-2}) for
the definition of $\cal A$), i.e. DP$({\cal A}) =: {\cal A}_kx^k/r$.
The above equation looks formally like a standard Newton equation. However,
it could not be interpreted as the Newton equation because its r.h.s. is not 
{\it a priori} given (it must be calculated from field equations).

To correctly interpret (\ref{Newton}) we have to take into account
(\ref{E-T}). Now, calculating $\ba$ in terms of $\bE^T_{(1)}$ and 
inserting into (\ref{Newton}) we obtain the following 
relation between $\bE^T_{(1)}$
and $\bD_{(-2)}$:
\beq   \label{dynamical}
   \mbox{DP} \left(   4r_e^4 (\bE^T_{(1)})^r -
\lambda_e (\bD_{(-2)})^r  \right) =0\ ,
\eeq
where  $r_e:= \sqrt{|e|/4\pi b}$ and $\lambda_e:= e^2/6\pi m$. $(\bF)^r$
denotes the radial component of a 3--vector $\bF$.
The main result of \cite{charge} consists in the following
\begin{TH}
Born--Infeld field equations supplemented by the dynamical condition
(\ref{dynamical}) define perfectly deterministic theory, i.e. initial
data for field and particle uniquely determine the entire evolution
of the system.
\end{TH}

\section{Duality invariance}

To show that the above theory can be generalized to include also Born--Infeld
dyons let us introduce a complex notation:
\beq
\bX &:=& \bD + i\bB\ ,\\
\bY &:=& \bE + i\bH\ .
\eeq
It was shown \cite{Gibbons} that the nonlinear electrodynamics is
invariant under duality rotations:
\beq    \label{rotations}
\bX \rightarrow e^{i\alpha}\bX\ ,\ \ \ \ \ \
\bY \rightarrow e^{i\alpha}\bY
\eeq
if and only if the following relation is satisfied
\beq    \label{invariance}
\mbox{Im}\, (\boX\bY)=0\ .
\eeq
In the case of Born--Infeld theory
 the constitutive relation reads:
\beq   \label{constitutive}
\bY(\bX,\boX) = \frac{1}{b^2R} \left[ \left(b^2 + \frac 12 \bX\boX\right)\bX
- \frac 12 \bX^2\boX \right]
\eeq
with duality invariant 
$R=\sqrt{ 1 + b^{-2}\bX\boX - \frac 14 b^{-4}(\bX \times \boX)^2}$.
One immediately shows that $\bY$ given by (\ref{constitutive})
satisfies (\ref{invariance}) (it turns out that the duality invariance
of the Born--Infeld theory was already observed by Schr\"odinger
\cite{Schrodinger}).

Let us consider a dyon carrying  electric and magnetic charges $e$
and $g$ respectively. Define the complex charge:
\beq
Q   :=  e + ig\ .
\eeq
Now, instead of (\ref{D-2}) we obviously have 
\beq  \label{F-2}
\bX_{(-2)} = \frac{Q{\cal A}}{4\pi}\frac{\br}{r}
\eeq
and instead of  relations (\ref{chain}) we have 
$\bX \sim r^{-2}$ and $\bY \sim r^{-1}$. To obtain the duality invariant
generalizations of (\ref{E-T}) and (\ref{dynamical}) we proceed as follows:
observe that $\oQ\bY$ is duality invariant ($\oQ$ stands for the complex
conjugation of $Q$). Therefore, its real and imaginary 
parts are also invariant. Using this invariance let as make the duality
rotation $Q' = e^{i\alpha}Q = e'$ (i.e. $g'=0$) and calculate
\beq
\mbox{Re}\, (\oQ\bY) =
e\bE + g\bH = e'\bE'\ .
\eeq
But in the rotated frame (i.e. $(e',0)$) we may use results of the 
previous section: formula (\ref{E-0}) implies
\beq  
 \mbox{Re}\, (\oQ\bY_{(0)}) = b|Q|\frac{\br}{r}
\eeq
and (\ref{E-T}) leads to
\beq
 \mbox{Re}\, (\oQ\bY^T_{(1)}) =
\frac {b}{4}|Q|
 \left( 3\mbox{\boldmath $a$} - 
r^{-2}(\mbox{\boldmath $a$} \mbox{\boldmath $r$})
\mbox{\boldmath $r$} \right)\ .
\eeq
Now, instead of (\ref{Newton}) we have duality invariant
\beq   \label{Newton-d}
ma_k = \frac{|Q|b}{3}{\cal A}_k\ ,
\eeq
and finally the duality invariant dynamical condition reads:
\beq   \label{dynamical-d}
   \mbox{DP} \left\{ \mbox{Re} \left[ \oQ \left(  
4r_0^4 (\bY^T_{(1)})^r -
\lambda_0 (\bX_{(-2)})^r  \right) \right] \right\} =0\ ,
\eeq
where  
\beq
r_0^4 &:=& r_e^4 + r_g^4 = \frac{e^2 + g^2}{(4\pi b)^2} =
\frac{Q\oQ}{(4\pi b)^2}\ ,\\
\lambda_0 &:=& \lambda_e + \lambda_g =  \frac{e^2 + g^2}{6\pi m}
= \frac{Q\oQ}{6\pi m}\ .
\eeq

\section{Canonical formulation}

Now we show that the duality invariant dynamical condition 
(\ref{dynamical-d}) may be derived from the mathematically well defined
variational principle. In the absence of a magnetic charge one could guess
that such a principle should be based on the following Lagrangian:
\beq  \label{total}
L_{total} = L_{field} + L_{particle} + L_{int}\ ,
\eeq
with $L_{field}$ given by (\ref{Lag-BI}), $L_{particle} = -m\sqrt{1-v^2}$
and $L_{int} = A_\mu j^\mu$. Varying $L_{total}$ with respect to $A_\mu$
one obviously gets field equations (\ref{field-eqs1})--(\ref{field-eqs2}).
The variation with respect to a particle's trajectory leads to the standard
Lorentz equation
\beq  \label{Lorentz}
ma^\mu = eF^{\mu\nu}u_\nu\ .
\eeq
However, despite the fact that $F^{\mu\nu}$ is bounded, it is not regular 
at the particle's  position and, therefore, the r.h.s. of (\ref{Lorentz})
is not well defined. This was already our motivation to find the 
mathematically well defined dynamical condition (\ref{dynamical}) which
replaces ill defined equations of motion (\ref{Lorentz}).

Observe that when $g\neq 0$ the situation is even worse. Actually, Lorentz
equation (\ref{Lorentz}) may be easily generalized to a duality
invariant formula
\beq  \label{Lorentz-d}
ma^\mu = (eF^{\mu\nu} + g{*G}^{\mu\nu})u_\nu\ ,
\eeq
but now both $F^{\mu\nu}$ and $G^{\mu\nu}$ are highly singular near a 
particle. 

It was shown in \cite{canonical} how to construct a consistent variational
principle for a pure electric monopole. Euler-Lagrange equations implied
by this variational principle are perfectly equivalent to the dynamical
condition (\ref{dynamical}). Moreover, it gives rise to a consistent
Hamiltonian structure with well defined Poisson bracket. It turns out that
this construction may be immediately generalized to include Born--Infeld dyon.
However, due to the fact that one uses a particle's rest frame the whole
procedure can not be in a straightforward way generalized to a many
particles case. In the present Letter we propose a new variational principle
which is valid for an arbitrary (finite) number of Born--Infeld dyons.
Moreover, it is much simpler that the one proposed in \cite{canonical}.

Let us consider $N$ Born--Infeld dyons: $(m_l,Q_l =e_l + ig_l);\ l=1,2,...,N$.
The energy of the composed "$N$ dyons + field" system in a fixed inertial
laboratory frame is given by:
\beq   \label{energy}
H = \sum_{i=1}^{N} \sqrt{m_i^2 + \bp_i^2} +
V(\bq_1,...,\bq_N)\ ,
\eeq
where $\bp_i = m\bv_i/\sqrt{1-\bv_i^2}$ denotes a "kinetic" momentum of an
$i$-th dyon and the function
\beq \label{V}
V(\bq_1,...,\bq_N) := \int_{\{\bq_1,...,\bq_N\}} d^3x\, T^{00}
\eeq
defines the  energy of the field configuartion $(\bX,\boX)$
($T^{00}$ denotes corresponding component of an energy-momentum
tensor of the Born--Infeld theory).
 The integral in (\ref{V}) is defined on a punctured 3-dimensional
(constant time) space where positions of $N$ dyons $\{\bq_1,...,\bq_N\}$ 
are excluded.   Moreover, in the phase
space of our system, parameterized by $(\bq_i,\bp_i)$ in a "dyons sector"
and by $(\bX,\boX)$ in a "field sector", define the following  
Poisson bracket:
\beq   \label{Poisson}
\{ {\cal F},{\cal G} \} & :=& \sum_{i=1}^{N} 
\frac{\partial{\cal F}}{\partial\bq_i} \cdot
\frac{\partial{\cal G}}{\partial\bp_i}
+ \frac 1i \int_{\{\bq_1,...,\bq_N\}} d^3x\, \left[
\frac{\delta{\cal F}}{\delta\bX}\cdot \nabla \times 
\frac{\delta{\cal G}}{\delta\boX} - (\bX \rightleftharpoons \boX) \right]
\nonumber\\ &-&
({\cal F} \rightleftharpoons
 {\cal G})\ ,
\eeq
for any two functionals $\cal F$ and $\cal G$. One may prove that the
formula (\ref{Poisson}) indeed defines a Poisson bracket (i.e. the Jacobi
identity is satisfied). With this 
definition we have the following "commutation relations" between dyons 
and fields variables:
\beq
\{ (\bq_i)^k,(\bp_j)_l\} &=& \delta_{ij}\delta^k_l\ ,\\
\{ X_k(\bx),\overline{X}_l(\by)\} &=& 2i \epsilon_{klm}\partial^m
\delta^3(\bx-\by)
\eeq
and remaining brackets vanish.

Let us treat $H$ given by (\ref{energy}) as the "dyons + fields"
Hamiltonian $H=H(\bq_i,\bp_i|\bX,\boX)$ and look for the 
corresponding Hamilton equations. In the "field sector" everything 
is clear:  fields equations
\beq
i\dot{{\bX}} = \{\bX,H\} = \nabla\times \bY\ 
\eeq
supplemented by the Gauss law $\nabla\cdot\bX=0$ are equivalent
to the Born--Infeld  field equations outside the dyons trajectories.
Observe, that $\bY$ is conjugated to $\bX$ {\it via}
\beq
\bY = \frac{\delta H}{\delta \bX} \ .
\eeq
 Now, in the
"dyon sector" one has obviously
\beq
\dot{{\bq}}_i = \{\bq_i,H\} = \bv_i\ .
\eeq
The only nontrivial thing is to evaluate
\beq   \label{dot-pi}
\dot{{\bp}}_i = \{\bp_i,H\} = - \frac{\partial}{\partial{\bq}_i}
V(\bq_1,...,\bq_N)\ .
\eeq
In the Appendix we show that (\ref{dot-pi})
is equivalent to
\beq    \label{Newton-d-i}
m_i(\ba_i)_k = \frac{|Q_i|b}{3}{\cal A}_{(i)k}\ ,
\eeq
where $\ba_i$ denotes the acceleration of an $i$--th dyon in its rest frame
and ${\cal A}_{(i)k}$ stands for a dipole part (its $k$--th component)
of the function ${\cal A}_{(i)}$ which defines the behaviour of $\bX_{(-2)}$
near an $i$--th dyon according to (\ref{F-2}).
Therefore, (\ref{Newton-d-i}) is equivalent to the $i$--th dynamical 
condition.  This way we have proved the following
\begin{TH}
The Hamiltonian (\ref{energy}) together with the Poisson bracket
(\ref{Poisson}) define the consistent canonical structure of a system
of $N$ Born--Infeld dyons.
\end{TH}
Let us observe that there is no "interaction term" in (\ref{energy}).
All information about the interaction between dyons and the field 
is encoded in the boundary conditions for the field variables
which have to be satisfied near dyons positions $\bq_i$, i.e. on the
multicomponent boundary of the punctured (constant time) space.
From the point of view of dyons dynamics the funcion (\ref{V}) plays
the role of a potential energy stored in the "field sector".

Obviously, performing the Legendre transformation in the "dyons sector"
one gets the corresponding Lagrange function
\beq   \label{lag}
L(\bq_i,{{\bv}}_i) = - 
\sum_{i=1}^{N} m_i\sqrt{1 - \bv_i^2} -
V(\bq_1,...,\bq_N)\ .
\eeq
\begin{TH}
The Euler--Lagrange equations implied by $L$
are equivalent to the $N$ dynamical conditions for dyons dynamics.
\end{TH}
The proof is straightforward. 
From the point of view of dyons dynamics the structure of (\ref{lag}) 
is evident: "kinetic energy -- potential energy". But in the "field sector"
(\ref{lag}) still generates the Hamiltonian dynamics because the field
generator is given by (\ref{V}). Therefore, (\ref{lag}) is a nice example of
a "mixed generator" called in the analytical mechanics a Routhian function.

\section{Concluding remarks}

Finally, let us make few remarks:

1. Let us observe that the force in a Newton--like equation
(\ref{Newton-d}) does not depend on a sign of $Q$ (contrary to the
Lorentz equation (\ref{Lorentz}) or (\ref{Lorentz-d})). It is a characteristic
feature of the self--interaction force already present in the Lorentz--Dirac
equation: $ma^\mu = eF^{\mu\nu}_{ext}u_\nu + \lambda_e(\dot{a}^\mu -
a^2u^\mu)$. The external force $eF^{\mu\nu}_{ext}u_\nu$ does depend on
a sign of $e$ but a self--force  proportional to $\lambda_e$ does not
($\lambda_e \sim e^2$).

2. The mass of dyon solution in the non--Abelian Yang--Mills--Higgs
theory in the BPS limit is given by
\beq  \label{BPS-mass}
M_{BPS} = a|Q|\ ,
\eeq
where $a$ stands for the vacuum expectation value of the Higgs field
(see \cite{Goddard}, \cite{BPS}, \cite{Olive}). Now, observe that the
l.h.s. of Newton--like equation (\ref{Newton}) contains purely mechanical 
quantities -- mass $m$ and acceleration $a^k$, whereas its r.h.s. contains
only electromagnetical quantities. The quantity $b|Q|/3$ looks formally
like a BPS mass with $a=b/3$. With this identification (\ref{Newton}) could be
rewritten in a suggestive form:
\beq
ma_k = M_{BPS}\,{\cal A}_k\ .
\eeq
Observe, that duality invariant Lorentz equation (\ref{Lorentz-d}) 
has no such a property.
Of course we do not claim that this identification has any 
fundamental meaning. However, our observation is supported by the fact that 
in string theory the $b$--parameter of
Born--Infeld action arises as a function of a vacuum expectation value
of a dilaton field \cite{string}.

3. The remarkable feature of the Hamiltonian (\ref{energy}) and Lagrangian
(\ref{lag}) is the absence of an interaction term. After removing the dyons
positions the nontrivial topology of the space ${\bf R}^3 - 
\{\bq_1,...,\bq_N\}$ requires very nontrivial boundary conditions for the
field variables at the multicomponent boundary  $\partial({\bf R}^3 - 
\{\bq_1,...,\bq_N\})$. Therefore, in a sense, the interaction is implied by a
space-time topology. Nevertheless, the above theory is not  of the topological
type (see e.g. \cite{Birmingham}), i.e. it is not true that its action does not
depend on a space--time metric.

4. It is interesting to note that the above feature is lost in the Maxwell
case. This is due to the fact that in this case the self--energy of a point
charged particle is infinite. To renormalize this theory one subtracts a
Coulomb term $\bX_{(-2)}$ (with ${\cal A}=1$)  which produces this infinity. 
But this subtraction leads to a mixed term  
in the energy functional of the form $\boX_{(-2)}\cdot(\bX - \bX_{(-2)})$. This
term is integrable (in a sense of the Cauchy principal value) and plays the
role of a gauge--invariant interaction term \cite{term}. Therefore, theorems
4 and 5 can not be applied to Maxwell electrodynamics. It turns out that these
theorems are connected with an analytical structure of classical (in general
nonlinear) electrodynamics of point--like objects. This point will be fully
clarified in the next paper.

5. As in the Maxwell theory the consistency with quantum mechanics implies
quantisation condition for dyons charges $Q_i$. It turns out that
the field dynamics does not play any role and the quantisation condition
is the same as in the Maxwell case. Using e.g. methods of \cite{Goddard}
one easily recovers duality invariant Zwanziger--Schwinger condition 
\cite{Z-S}:
\beq
\mbox{Im} (\overline{Q_i}Q_j) = 2\pi \hbar n_{ij}
\eeq
with $n_{ij}$ integers.

\section*{Appendix}
\setcounter{equation}{0}
\def\theequation{A.\arabic{equation}}

To prove the equivalence of (\ref{dot-pi}) and (\ref{Newton-d-i})
let us observe that for any field functional of the form
\beq
{\cal F}(\bq_1,...,\bq_N) = \int_{\{\bq_1,...,\bq_N\}} f(\bX,\boX)\,d^3x\ ,
\eeq
we have
\beq
\frac{\partial}{\partial(\bq_i)^k} {\cal F}(\bq_1,...,\bq_N) = 
- \{ {\cal F}(\bq_1,...,\bq_N),{\cal P}_k(\bq_i)\}\ ,
\eeq
where
\beq
{\cal P}_k(\bq_i) := \int_{\{\bq_i\}} T^0_{\ \, k} \,d^3x
\eeq
denotes $k-$th component of field momentum outside the $i-$th dyon.
This relation follows from the Poincar\'e algebra structure.
Using (\ref{Poisson}) it is easy to prove that
\beq
\{ V(\bq_1,...,\bq_N),{\cal P}_k(\bq_i)\} = \int_{\partial\Sigma_i} 
T^{00}n_k\, d\sigma\ ,
\eeq
where ${\partial\Sigma_i}$ denotes a boundary of $\Sigma_i := {\bf R}^3 -
\{\bq_i\}$, $\ n_k$ stands for a unit normal to ${\partial\Sigma_i}$ and
$d\sigma$ denotes its standard measure. 

Now,  $\partial\Sigma_i$ has two components.
The boundary integral over the component at infinity vanishes but the one
over the component surrounding $\{\bq_i\}$ 
gives exactly $|Q|b{\cal A}_k/3$ (see \cite{charge}).

\section*{Acknowledgements}

One of us (D.C.)  thanks Alexander von Humboldt--Stiftung for 
the financial support.

\end{document}